\documentclass[pra,aps,showpacs,epsf,twocolumn]{revtex4}

\usepackage{graphicx}
\usepackage{epsfig}
\usepackage{color}

\begin{document}

\title{Formation of fermionic molecules via interisotope Feshbach resonances}
\author{E.G.M. v. Kempen, B. Marcelis, and S.J.J.M.F. Kokkelmans}
\affiliation{Eindhoven University of Technology, P.O.~Box~513, 5600~MB  Eindhoven, The Netherlands}
\date{\today}

\begin{abstract}
We perform an analysis of recent experimental measurements and improve the lithium interaction potentials. For $^6$Li a consistent description can be given. We discuss theoretical uncertainties for the position of the wide $^6$Li Feshbach resonance, and we present an analytic scattering model for this resonance, based on the inclusion of a field-dependent virtual open-channel state. We predict new Feshbach resonances for the $^6$Li-$^7$Li system, and their importance for different types of crossover superfluidity models is discussed.
\end{abstract}

\pacs{34.50.-s, 02.30.Mv, 03.75.Nt, 11.55.Bq}

\maketitle

Resonances in cold atomic gases offer the key to connections with
challenging condensed matter physics. In particular, resonances
make lithium atomic systems very versatile. The first
Bose-Einstein condensates (BEC) were quite small in number due to a
negative scattering length~\cite{bradley}, later Feshbach
resonances~\cite{feshbach} have been used to create condensates
with positive scattering length and to generate bright
solitons~\cite{salomon,hulet2} by changing the scattering length
$a$ back to negative. Even more interesting is the usability to
form molecules, since the Feshbach resonance results from bringing
a molecular state on threshold. The connection between fermionic
atoms and composite bosons (molecules) has great impact for the
study of the well-known crossover problem between BEC and
Bardeen-Cooper-Schrieffer (BCS)-type superfluidity\cite{nozieres,jin,Zwierlein04}.

In this paper, we study several Feshbach resonances in the lithium
system. We first review the knowledge of the interatomic
interaction potentials, and use experimental data as input to
improve these potentials. We discuss the special situation
for the wide $^6$Li resonance, where the background scattering length depends strongly on the magnetic field. This will
be interpreted as a field-dependent virtual state (a second
resonance) which is situated close to threshold. The full
energy-dependent scattering process can be parametrized according
to a simple analytical model that encapsulates both
field-dependent resonances. Further, we apply our knowledge of the
lithium interactions to a system of $^6$Li-$^7$Li, and find
several Feshbach resonances which are accessible in current
experimental setups. The underlying molecular state is of a
composite fermionic nature, which allows for a new type of
crossover physics --- the transition of an atomic BEC to a
molecular Fermi-type of superfluidity. Recently, Feshbach resonances in a heteronuclear Bose-Fermi mixture have been observed~\cite{stan,inouye}, where polar fermionic molecules underly the resonance state.

\begin{figure}[t]
    \includegraphics[width=\columnwidth]{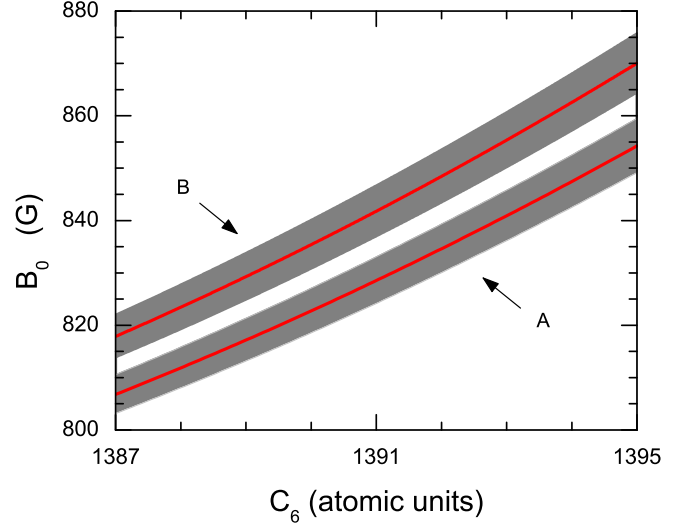}
    \caption{(color online) Dependency of the wide-resonance position $B_{0}$ on $C_6$, according to the second analysis (see text). The shaded area indicates the associated uncertainty. The curve indicated with A (B) is obtained making use of the exchange energy according to~\cite{Zemke99} (\cite{Marinescu96}).} \label{B0_vs_C6}
\end{figure}

For an accurate prediction of resonance properties, we need a detailed understanding of the actual interatomic potentials. Here we describe how we improved the precision of existing potentials by using recent experimental measurements as input. The potentials can be divided in two radial intervals. For large interatomic separations $r$ the potential is given by the sum of the dispersive Van der Waals tail $V^{\rm vdw}(r)=-C_6/r^6 -C_8/r^8
-C_{10}/r^{10}$ and the exchange contribution $V^{\rm ex}_S(r)=(-1)^{S+1} C^{ex} r^{7/2\alpha -1}e^{-2 \alpha r}$~\cite{Smirnov65}, resulting in two potentials: a singlet ($S=0$) and a triplet ($S=1$) potential.
The coefficient $C^{ex}$ is taken from Refs.~\cite{Zemke99,Marinescu96}, $\alpha$ is directly related to the ionization energy
$\alpha ^2/2$~\cite{NISTwww}, and $C_8$ and $C_{10}$ are taken from Refs.~\cite{Yan96}. For smaller $r$ we 
use the model singlet and triplet potentials which have also been used in Refs.~\cite{Moerdijk94,vanAbeelen97}.

These short-range and long-range potentials are smoothly connected
at $r=18a_0$, with $a_0$ the Bohr radius. 
To overcome the inaccuracies of the short range potentials, we make use of the accumulated phase method~\cite{Moerdijk94}. A boundary
condition is applied on the partial-wave radial wave functions at
$r=17.5a_0$ in the form of a WKB phase
$\phi_{S,T}(E,\ell)=\phi^0_{S,T}(E,\ell)+\Delta \phi_{S,T}$. The first
term on the right is calculated by radial integration of the model
potential up to $17.5a_0$ and is expected to account for the
energy and angular momentum dependence of the accumulated phase to
a sufficient degree of accuracy. The second term is an energy and
angular momentum independent shift of the phase, determined from experimental data. These corrections $\Delta \phi_{S,T}$ to the accumulated singlet and triplet phases can be converted to the more physical quantities $\nu_{DS,DT}$, which are the fractional vibrational quantum numbers at dissociation.

We determine the free parameters of our interaction potentials $\nu_{DS}$, $\nu_{DT}$, and $C_6$ from experimental input by means of a $\chi ^2$ minimization.
An inter-isotope analysis in which $^7$Li is related to $^6$Li by means of a simple mass-scaling relation failed, yielding inconsistent results for $\nu_{DS}$. This is a strong indication of a breakdown of the Born-Oppenheimer approximation for the singlet potential. Such a breakdown was demonstrated in detailed spectroscopy~\cite{Wang02}. We therefore avoid mass scaling of the singlet potential, and we perform two different analyses. In the first analysis, we only take $^6$Li data into account. In the second analysis we investigate $^7$Li as well, however, we only do a mass scaling for the triplet potential. Our total set of $^6$Li experiments comprises 6 data points. The zero-crossing of the
scattering length of a system in the two lowest hyperfine states~\cite{OHara02,Jochim02}. In the same spin state configuration,
the positions of the narrow~\cite{Strecker03} and wide~\cite{Zwierlein04} Feshbach resonances. The measurement of the scattering length in the lowest and third to lowest hyperfine state~\cite{OHara00} and the binding energy of the most weakly
bound triplet state~\cite{Abraham97}. 

\begin{figure}[t]
    \includegraphics[width=\columnwidth]{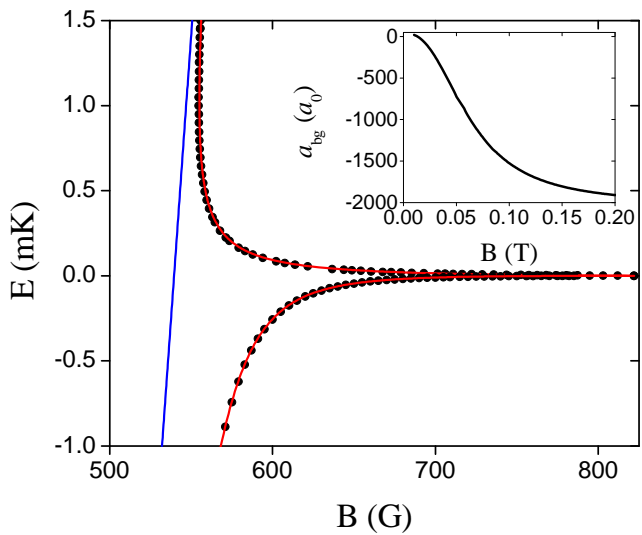}
    \caption{(color online) Energy of the dressed \mbox{(quasi-)}molecular state as obtained with a full coupled-channels calculation (dots), with respect to the $|f_1,m_{f_1}\rangle \otimes |f_2,m_{f_2}\rangle =1/2,1/2\rangle \otimes 1/2,-1/2\rangle$ threshold. The straight line is the bare closed-channel energy $\epsilon_b(B)$, and the energy of the dressed state according to our model is indicated by the curved line. Inset: corresponding background value of the scattering length as a function of the magnetic field $B$.} \label{abg}
\end{figure}

In our first analysis we obtain a minimum in the reduced $\chi^2$ distribution of $\chi ^2=0.5$. The corresponding parameter values are 
$\nu_{DS}=0.3496(5)$, equivalent with a singlet scattering length
$a_S=45.3(1)a_0$ and $\nu_{DT}=0.9954(2)$, corresponding to
$a_T=-2025(70)a_0$. For the leading dispersion coefficient we find
$C_6=1388(6)$ at.~units. This result has been obtained with a $C^{ex}$ of Ref.~\cite{Marinescu96}. When we weaken $C^{ex}$ to the value of Ref.~\cite{Zemke99}, we find that the optimal $C_6$ is shifted to $C_6=1390$ at.~units.
The scattering lengths found are consistent with previous
determinations~\cite{Abraham97,vanAbeelen97}. Our $C_6$
coefficient agrees with the values found in {\it ab
initio} calculations~\cite{Yan96,Derevianko01}.

The objective of a second analysis~\cite{SecAnalysis} is to reevaluate the position
of the wide s-wave Feshbach resonance of $^6$Li in the two lowest
hyperfine states, without making use of the experimental result $B_0$=822G~\cite{Zwierlein04}. Hereto we want to combine all available cold collision data on lithium, and we add the positions of three p-wave Feshbach resonances, which have been measured
recently~\cite{Zhang04}, and experimental data of $^7$Li~\cite{Abraham95,salomon}, to the set of experiments. In this combined isotope analysis we
perform a mass scaling procedure for the triplet boundary
condition only. As explained above, we will not mass-scale the singlet potential but rather optimize the boundary conditions for the singlet potential independently for the two isotopes, by making $\nu_{DS,7}$ a free parameter independent of $\nu_{DS,6}$.

Since the p-wave resonances are measured with high accuracy, we also allow for small corrections of the angular momentum dependence of $\phi^0_{S}(E,\ell)$ via the parameter $\Delta \phi_S^l$ by means of an addition $\Delta \phi_S^{l} \cdot  l(l+1)$ cf.~\cite{kempen02}.
By optimizing the interaction parameters
($\nu_{DS,6}$, $\nu_{DS,7}$, $\nu_{DT,6}$, $\Delta \phi_S^l$) for various fixed
values of $C_6$ we are able to obtain a minimum reduced $\chi ^2$
of ~0.7. The dependency of $\chi ^2$ on $C_6$ is rather weak and therefore the
set of experiments does not restrict the $C_6$ coefficient to an
acceptable degree. However, the minimal $\chi ^2$ occurs for $C_6=1390.6$ at.~units close to {\it ab initio} values, when using $C^{ex}$ from Ref.~\cite{Zemke99}, positioning the wide Feshbach resonance at $B_{0}=826$~G. 
We estimate $808$~G $<B_{0}<846$~G for $1388<C_6<1393$ at.~units, see Fig.~\ref{B0_vs_C6}. For $C^{ex}$ from Ref.~\cite{Marinescu96} we find a minimum $\chi ^2=0.5$ for $C_6=1395.6$.

From now on, the properties of the Feshbach resonances will be derived from our first analysis. First, we study the wide $B_0 = 822$~G Feshbach resonance in the two lowest hyperfine states of $^6$Li. This resonance is quite remarkable for two reasons: it has a large width of the order of 100~G, and its background scattering length $a_{\rm bg}$ is strongly depending on the magnetic field, which can be seen from the inset of Fig.~\ref{abg}. At zero field, $a_{\rm bg}$ is 3 and positive, while for large field values $a_{\rm bg}$ is large and negative, indicating the presence of a nearby virtual
state in the open-channel subspace $P$~\cite{Marcelis04}.
Consequently several important quantities, such as the $S$ and
$T$ matrices which summarize the collision process, depend non-trivially on the collision energy $E$. We will here apply the model discussed in Ref.~\cite{Marcelis04} to
this s-wave Feshbach resonance. This model takes the virtual state
into account explicitly, and gives an analytical description of
all important two-body quantities near the Feshbach resonance.

In general, the relation between the
background scattering length $a_{\rm bg}$, the range
of the potential $a_{\rm bg}^P$, and the virtual state pole
$\kappa_{\rm vs}$, is given by $a_{\rm bg} = a_{\rm bg}^P -
1/\kappa_{\rm vs}$. The range of the potential is related to the
Van der Waals coefficient $C_6$, and does not depend on the
magnetic field. Therefore, we account for the field dependence of
$a_{\rm bg}$ by generalizing the model of Ref.~\cite{Marcelis04}
to the case of a field-dependent virtual-state $\kappa_{\rm
vs}(B)$. The complex energy shift is then given by
\begin{eqnarray}
    A(E,B) &=& \Delta_{\rm res}(E,B)-\frac{i}{2}\Gamma(E,B) \nonumber\\
    &=& \frac{-iA_{\rm vs}(B)}{2 \kappa_{\rm vs}(B)[k+i\kappa_{\rm vs}(B)]},
\end{eqnarray}
where $A_{\rm vs}(B)$ is related to the coupling matrix element
between the open-channel virtual state and the closed-channel
bound state responsible for the Feshbach resonance. Our wavenumber units are such that
$E=k^2$.

The total scattering length is then given by
\begin{equation}
    a(B) = a_{\rm bg}^P - \frac{1}{\kappa_{\rm vs}(B)} - \lim_{E \rightarrow 0}\frac{\Gamma(E,B)/2}{k[\epsilon_b(B) + \Delta_{\rm res}(E,B)]}, \label{aofB}
\end{equation}
where $\epsilon_b(B) = \Delta \mu^{\rm mag}(B-\overline{B}_0)$ is the energy of the bare closed-channel bound state, $\Delta \mu^{\rm mag}=2.0 \mu_B$ the magnetic moment, $\mu_B$ the Bohr magneton, and $\overline{B}_0 = 539.5$~G the field where the bare closed-channel energy crosses threshold. For a fixed $B$ value, the
parameters $A_{\rm vs}(B)$ and $\kappa_{\rm vs}(B)$ are obtained
by fitting Eq.~(\ref{aofB}) to the coupled-channels result for
$a(B)$, using two close-lying data points where we assume $A_{\rm vs}$ and $\kappa_{\rm vs}$ locally constant. Repeating this for
every field value, we obtain explicit expressions for the
non-trivial energy dependence of the complex energy shift, using
only the zero-energy information contained in $a(B)$. Note that
there are only two free parameters, and $a_{\rm bg}(B)$ is fixed
once $\kappa_{\rm vs}(B)$ is known.

The fit functions are summarized in
Table~\ref{table}. In Fig.~\ref{abg} we compare the dressed \mbox{(quasi-)}molecular state calculated by coupled-channels methods and by our analytical model, which agrees excellent. Therefore this model can be used to analytically describe the $S$ and $T$ matrices~\cite{Marcelis04}, scattering
phase-shifts, etc., with similar precision as a full coupled channels calculation, for a large range of energies and magnetic fields. We note that also the narrow 543~G Feshbach resonance (width of order 0.1~G) can be described by Feshbach theory. Here, however, an easier description is possible based on only one single background part and a single resonance state, since the narrow resonance has a `local' background scattering length of the order of $a_{\rm bg}^P$. For narrow resonances~\cite{depalo,petrov2}, the typical resonance features in the continuum scattering which depend on the details of the potential, are very important to the BEC-BCS crossover physics, since they are visible for energies less than the Fermi energy, resulting in a non-universal crossover picture.

\begin{figure}[t]
    \includegraphics[width=\columnwidth]{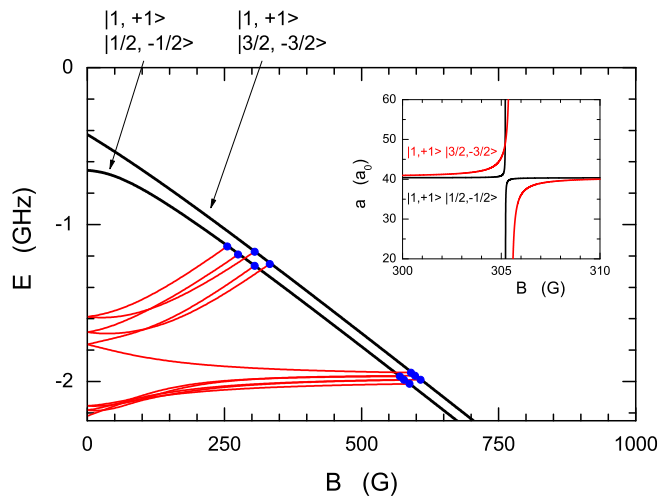}
    \caption{(color online) Two thresholds for the $^6$Li-$^7$Li system which show the coincidence of two Feshbach resonances. Inset: scattering length as a function of magnetic field, for the two coinciding Feshbach resonances.} 
    \label{thresholds}
\end{figure}

\begin{table}
\caption{Parameters of the virtual-state model. For the magnetic
fields of interest $(530$~G$\lesssim B \lesssim 830$~G), the
field dependent parameters are given by a third-order polynomial
fit, $c_0 + c_1 B + c_2 B^2 + c_3 B^3$. For $A_{\rm vs}(B)$ the
units are given by $[c_n] = K^2 G^{-n}$, and for $\kappa_{\rm
vs}(B)$ the units are given by $[c_n] = K^{1/2} G^{-n}$. The range
of the potential is given by $a_{\rm bg}^P = 45$~a$_0$.}
\label{table}
\begin{center}
\begin{tabular}{|l|l|l|l|l|}
\hline & $c_0$ & $c_1$ & $c_2$ & $c_3$ \\ \hline $A_{\rm vs}(B)$ &
$1.62\cdot 10^{-5}$ & $-5.10\cdot 10^{-8}$ & $6.07\cdot 10^{-11}$ &
$- 2.56\cdot 10^{-14}$ \\ \hline $\kappa_{\rm vs}(B)$ & $4.65\cdot
10^{-2}$ & $- 1.54\cdot 10^{-4}$ & $1.92\cdot 10^{-7}$ & $- 8.26\cdot 10^{-11}$ \\ \hline
\end{tabular}
\end{center}
\end{table}

Now we turn to the final topic of this paper, the study of Feshbach resonances in a $^6$Li fermion - $^7$Li boson mixture. We use again our first analysis of the interaction parameters for $^6$Li, and perform a simple mass scaling of the accumulated phases for the $^6$Li-$^7$Li system. As discussed before, this leads to less accurate predictions for the resonance positions as suggested by the accuracies from the $^6$Li interaction parameters. We estimate the inaccuracies in the mixed isotope resonance positions to be of order a few Gauss, due to inaccuracies of the mass-scaling relations.

Feshbach resonances in the $^6$Li-$^7$Li system have been studied before in Ref.~\cite{vanAbeelen97}, where only the case of magnetically trappable atoms was investigated. Moreover, those resonances are accompanied by large inelastic exchange losses. We investigate only $^6$Li-$^7$Li hyperfine state combinations where exchange losses are absent. Within these boundaries, there are still numerous resonances present, and we restrict ourselves to the most interesting results. The $^6$Li-$^7$Li $|1/2,1/2\rangle \otimes |1,1\rangle$ channel has the lowest energy in the two-body hyperfine diagram. Therefore, this channel will not suffer from magnetic dipolar relaxation. We find five Feshbach resonances at magnetic field values of 218~G, 230~G, 251~G, 551~G, and 559~G.
Measurements of these resonances might provide the missing information to exactly locate the position of the wide $^6$Li resonance, as all mixed resonances arise from the same underlying bound state in the triplet potential.

Feshbach resonances between bosons and fermions give rise to a fundamentally different type of crossover physics. Already some work on the interactions in Bose-Fermi mixtures can be found in the literature~\cite{viverit,mackie}, however, in order to describe correctly the many-body physics and interactions close to resonance, more research is needed. To make the system feasible, some requirements have to be fulfilled. For a stable BEC, a positive scattering length for the bosons is required. Approaching the resonance from the atomic side, the mixed boson-fermion scattering length will become negative, and stability of the system could become and issue. At the other side of the resonance, it is important that the effective interaction between the fermionic molecules is attractive. Point-like composite fermions do not undergo s-wave collisions. However, close to resonance the molecules are long stretched, and an effective interaction mediated via the bosons could be possible. An approach similar to Ref.~\cite{petrov} could be conclusive on this and on the expected dependence of the inelastic rate coefficient on the scattering length. Also, the effect of Pauli blocking will not be as strong as in Ref.~\cite{petrov} since a three-body decay process with only one fermion and two bosons involved is possible, however, still a reduction with respect to the pure bosonic case could be expected. 

Another interesting situation occurs when two Feshbach resonances coincide. Further away from resonance, where the size of the molecule is comparable to the size of the potential, s-wave collisions are not allowed for these composite fermions. Therefore, in order to preserve superfluid behavior in this region, two different molecular spin configurations are needed to allow for s-wave collisions between molecules. Moreover, it is desirable that the two resonances responsible for the molecule formation coincide. Such a coincidence can be found from Fig.~\ref{thresholds}, where the  $^6$Li-$^7$Li $|1/2,-1/2\rangle \otimes |1,1\rangle$ and $|3/2,-3/2\rangle \otimes |1,1\rangle$ bound states and thresholds are plotted as a function of magnetic field. Every crossing of a bound state with threshold indicates the position of a Feshbach resonance. It can be seen that at $B=305$~G two Feshbach resonances coincide. These two threshold channels have the same bosonic $^7$Li state, but a different fermionic $^6$Li state. This coincidence is systematic, and will not depend on the interaction parameters. The scattering lengths as a function of the magnetic field can be seen in the inset of Fig.~\ref{thresholds}.

In conclusion, we analysed recent experimental measurements for lithium, and showed that mass scaling between $^6$Li and $^7$Li fails for the singlet potential. We investigated uncertainties in the wide $^6$Li resonance position, and demonstated an analytical model for this resonance that includes the nearby virtual state. Finally, we showed that $^6$Li-$^7$Li mixtures feature accessible Feshbach resonances, giving rise to fermionic molecules, yielding new BCS-BEC crossover physics.
  
We acknowledge discussions with B. Verhaar, M. Holland, J. Zhang, T. Bourdel, F. Chevy and C. Salomon, and we acknowledge the Netherlands Organization for Scientific Research (NWO). E.~K.~acknowledges support from the Stichting FOM, financially supported by NWO.


\begin{thebibliography}{99}
\bibitem{SecAnalysis} Here $C^{ex}$ is taken from ref.~\cite{Zemke99}.
\end{thebibliography}
\end{document}